\documentclass[12pt]{article}

\usepackage{epsfig}
\usepackage{graphicx}
\usepackage{graphics}
\usepackage{amssymb}
\usepackage{amsmath}
\usepackage{color}
\usepackage{textcomp}
\usepackage{latexsym}
\usepackage{cite}

\begin{document}

\begin{center}
{\Large{\bf Granular Fluids with Solid Friction and Heating}} \\
\ \\
\ \\
by \\
Prasenjit Das$^1$, Sanjay Puri$^1$ and Moshe Schwartz$^{2,3}$\\

$^1$School of Physical Sciences, Jawaharlal Nehru University, New Delhi 110067, India. \\
$^2$Beverly and Raymond Sackler School of Physics and Astronomy, Tel Aviv University, Ramat Aviv 69934, Israel. \\
$^3$Faculty of Engineering, Holon Institute of Technology, Golomb 52, Holon 5810201, Holon, Israel.
\end{center}

\begin{abstract}
We perform large-scale molecular dynamics simulations to study heated granular fluids in three dimensions. Granular particles dissipate their kinetic energy due to solid frictional interaction with other particles. The velocity of each particle is perturbed by a uniformly-distributed random noise, which mimics the heating. At the early stage of evolution, the kinetic energy of the system decays with time and reaches a steady state at a later stage. The velocity distribution in the steady state shows a non-Gaussian distribution. This has been characterized by using the Sonine polynomial expansion for a wide range of densities. Particles show diffusive motion for densities below the {\it jamming density} $\phi_{\rm J}$.
\end{abstract}

\newpage
\section{Introduction}\label{sec1}

In the past few decades, the study of granular materials has become attractive because of the novel properties discovered and the many industrial applications. Granular materials belong to a special state of matter. They have properties similar to both solids and liquids~\cite{jn96,dg99,at06}. For example, the flow of sand through the narrow neck of an hourglass certainly looks like a flow of a liquid. Nevertheless, one can walk on the sandy surface of a desert which supports the weight of the walker like solids do. Also, a stable sand pile acts as a solid when stress is applied along the vertical direction and deforms like a liquid when force is applied along horizontal directions. The constituting elements of such materials are solid particles, which are polydispersed in size, ranging from a few microns to few centimeters and also polydispersed in shape, often modeled by spheres, cylinders or needles~\cite{jd94,bp04}. The particles in a granular system have rough surfaces and are subjected to nonconservative contact forces such as solid friction~\cite{dps16,bes10,sb11}, cohesion~\cite{pl01}, and inelastic collisions~\cite{dp03,sdsp03,ap03,ap07} which can be described by the coefficient of restitution less than unity. Because of the dissipative nature of particle-particle interactions, the kinetic energy of relative motion of particles decreases with time~\cite{dps16,ap03,ph83} which is in contrast to molecular gases, where the interaction among the particles conserves kinetic energy. Various interesting phenomena have been observed due to the continuous dissipation of energy, such as clustering, pattern formation~\cite{dps16,dp03,sdsp03,ap03,ap07}, and inelastic collapse~\cite{gz93,gtz93}, etc. The detailed understanding of the dynamical properties of granular gases helps us to explain the early stages of planet formation or the planetary rings. Recently, Brilliantov \textit{et al.}~\cite{bkb15} described the size distribution of particles in Saturn's rings from the concept of granular aggregation and fragmentation. Naturally, the behavior of granular matter is also important in the field of geophysics, e.g., to describe avalanches, debris flow, and mud flow, etc.~\cite{pgp03,gg14}.

In experimental granular systems, the dissipative nature of particle-particle interactions and the athermal properties of particles contribute to many unique properties. In the absence of any external drive, the constituents of a granular system dissipate their kinetic energy, and the system becomes sluggish as time advances~\cite{bp04,dps16}. If energy is injected into a granular system to compensate for its losses in dissipative collisions, the system settles into a nonequilibrium steady state~\cite{ms00,ne98}. For example, in the steady state structure of planetary rings, the energy lost due to inelastic particle-particle collision is compensated by the input energy from gravitational interactions~\cite{sor09,bhl84}. A fascinating class of phenomena observed in driven granular systems, e.g., fluidization, convection, clustering, density wave formation, and fluctuations, etc~\cite{jn96,dg99,at06}. For example, the flow of granular matter through a hopper shows jamming, and the granular matter turns into a disordered solid~\cite{gm04}. However, in the presence of vibration, the granular matter continues to flow through the hopper. Granular materials also exhibit a broad range of pattern forming behaviors when excited by vibrating the container. Experimentally, depending upon the direction, amplitude and the frequency of vibration, formation of stripes, squares, hexagons, spirals, interfaces, and localized oscillons have been observed~\cite{mus95,ums96,gr00}. Thus, granular materials under excitation can be thought of as an example of a complex system. Also, granular mixtures segregate or unmix according to their sizes and masses in a rotating drum~\cite{zls94,ph99}. There are different ways to energize a given system: via shearing~\cite{bls01}, vibration~\cite{wp02,fm02}, or rotating walls of a container~\cite{zls94,ph99}, applying electrostatic~\cite{ao02} or magnetic forces~\cite{sak05}, etc. 

Heated granular gases have been studied extensively through analytical calculations~\cite{ne98,spe99,bt98,pmp02}, and computer simulations~\cite{ms98,po98,net99,kh04}. In simulations, thermostats are used to inject energy into a system. A few types of thermostats have been proposed in the literature. White-noise thermostats (WNT) are most commonly used, where all the particles are heated uniformly and independently by perturbing their velocity. Williams \textit{et al.}~\cite{wm96} studied the problem of heated granular gases using WNT. They modeled the dissipative interaction among the granular particles by inelastic collisions with the coefficient of restitution $e$ less than unity. When $e = 1$, the motion of particles are uncorrelated, and as $e$ decreases, long-range spatial correlations between particles develop. Murayama and Sano~\cite{ms98} have studied the velocity distribution  function (VDF) in two-dimensional vibrated bed of granular material using a viscoelastic particle model. Kawarada and Hayakawa~\cite{kh04} have also studied dynamical properties of granular materials in horizontally as well as vertically vibrated plane. In all the cases, the velocity follows a non-Gaussian distribution. Peng and Ohta~\cite{po98} used event-driven molecular dynamics (EDMD) simulation to study the dynamics in a heated hard-sphere system. In their study, density and velocity correlation functions are long-ranged and follow power-law scaling. Many computer simulations and experiments reported results showing VDFs that obey Gaussian behavior at small argument with exponential tails, where the transition from Gaussian to exponential depends on the density. Stretched exponential and even power-law functions~\cite{bp04,pl01} have also been observed. Recently, Bodrova~\textit{et al.}~\cite{bdp12} studied the velocity distribution in free as well as heated viscoelastic granular gases. They modeled the system using the hard sphere exclusion with velocity-dependent coefficient of restitution $e(v_{rel})$. In their study, the VDF is also a non-Gaussian function.

Most studies of granular materials consider two-body inelastic collisions as the mechanism of dissipation. This is reasonable for granular gases. However, in dense granular systems, the concept of binary collisions becomes less useful as each particle interacts simultaneously with many others, and solid friction plays an important role in the dissipation process~\cite{dps16}. In this paper, we study granular fluids whose energy is only dissipated by frictional interactions. In this context, we investigate granular fluids at low and high densities, both with and without heating by a WNT. Our primary goal in this paper is to study phenomenology in frictional powders, and compare it with results for the well-studied case with inelastic collisions.

Given this background, this paper is organized as follows. In Sec.~\ref{sec2}, we describe the details of modeling and velocity distributions. Our numerical results are described in Sec.~\ref{sec3}. Finally, we summarize our results in Sec.~\ref{sec4}.

\section{Modeling and Velocity Distribution Functions}\label{sec2}

\subsection{Details of Model}\label{sec2a}

We use standard molecular dynamics (MD) techniques for our simulations, where all the identical particles in the system are considered to be spherical in shape and of mass $m$. Any two particles with position vectors $\vec r_i$ and $\vec r_j$ interact via a two-body potential with a hard core of radius $R_1$ and a thin shell repulsive potential~\cite{dps16}. To be specific, we choose the potential to be of the following form:
\begin{eqnarray}
\label{soft}
V(r) &=& \infty, \quad r < R_1, \nonumber \\
&=& V_0 \left(\frac{R_2-r}{r-R_1}\right)^2, \quad R_1 \le r < R_2, \nonumber \\
&=& 0, \quad r \ge R_2,
\end{eqnarray}
where $r=\mid\vec r_i - \vec r_j\mid$ is the separation between the two particles, $V_0$ is the amplitude of the potential and $R_2 - R_1 < R_1$. Here, Eq.~(\ref{soft}) corresponds to a repulsive potential that rises steeply from zero at the outer boundary of the shell of radius $R_2$ to infinity at the hard core ($r=R_1$). The normal force applied by particle $i$ on particle $j$ is given by
\begin{eqnarray}
 \label{fn}
 \vec F_{ij}^n(r) = -\vec\triangledown_j V(r),
\end{eqnarray}
where $\vec\triangledown_j$ is the gradient with respect to $\vec r_j$. The corresponding solid friction force is given by
\begin{eqnarray}
 \label{ff}
 \vec F_{ij}^f(r) = \mu \mid \vec F_{ij}^n\mid \frac{\vec v_i - \vec v_j}{\mid \vec v_i - \vec v_j \mid},
\end{eqnarray}
where $\vec v_i$ and $\vec v_j$ are the linear velocities of particles $i$ and $j$ respectively. In Eq.~(\ref{ff}), $\vec F_{ij}^f$ only depends upon the direction of the relative velocity and not its magnitude. The tangential component dissipates energy in the sliding motion or rubbing of two grains. For simplicity, we did not consider rotational motion of the grains. In Eq.~(\ref{ff}), the relative velocity, and thereby the frictional force, has both tangential and normal components at the point of contact. In principle, it is easy to remove the normal component but this would have burdened the numerical calculations. Fortunately, the normal component does not play a significant role due to the stiffness of the radial potential. Our results are unchanged if we explicitly drop the normal component, as is the case for Coulombic friction. Further, Eq.~(\ref{ff}) reduces to Coulombic friction when the thickness of the thin repulsive shell tends to zero, i.e., effectively our model reduces to a hard-sphere model where the velocity difference cannot have a normal component at the contact. Thus, the frictional force becomes perpendicular to the normal force.

We use the following units for various relevant quantities: lengths are expressed in units of $R_1$, energy in $u=V_0/10$, temperature in $u/k_{\rm B}$, and time in $t_0=\sqrt{mR_1^2/V_0}$. For the sake of convenience and numerical stability, we set $R_1 = 1$, $R_2 = 1.1R_1$, $V_0 = 10$, $k_{\rm B} = 1$, and $m = 1$. Therefore, the time unit is $t_0=1/\sqrt{10}$ and this allows us to take moderately large $\Delta t$ in the simulation. The velocity Verlet algorithm~\cite{at87,fs02,dr04} is implemented to update positions and velocities in the MD simulations, and we used the integration time step $\Delta t=0.0005$ in scaled units. The total number of particles in the system is $N=350000$. To obtain the desired volume fraction, $\phi$, we vary the linear size of the system $L$. Periodic boundary conditions are employed in all directions.

To heat the system uniformly, we keep it in contact with a heat bath, which is modeled by the WNT mentioned earlier, i.e., independent white noise random forces acting separately on each particle \cite{ne98}. In practice this may be the air affecting the sand particles in a sand storm, external vibrations applied to the granular system to avoid plugs, etc. In this case, the equation of motion for the $j^{\rm th}$ particle can be written as
\begin{eqnarray}
 \label{heat}
 m\frac{d\vec v_j}{dt} = \vec F_j^{\rm tot} + \vec\eta_j,
\end{eqnarray}
where
\begin{eqnarray}
\label{net_f}
\vec F_j^{\rm tot} = \sum_{i\ne j}\left(\vec F_{ij}^{\rm n} + \vec F_{ij}^{\rm f}\right),
\end{eqnarray}
and $\vec\eta_j$ is the external stochastic force modeled to obey
\begin{eqnarray}
\label{noise1}
\langle \vec\eta_j\rangle &=& 0,\\
\label{noise2}
\langle \eta_{i,\alpha}(t) \eta_{j,\beta}(t')\rangle&=&m^2\xi^2\delta_{ij}\delta_{\alpha\beta}\delta(t-t'),
\end{eqnarray}
where $\alpha, \beta = x, y, z$ and $\xi$ characterizes the amplitude of the stochastic force.

Before proceeding, we remark that several other thermostats have also been employed in the literature, e.g., the {\it Gaussian thermostat} \cite{as03} and the {\it Langevin thermostat} \cite{hh03,crg15}. In certain physical situations, these may constitute more realistic models for heated powders. Here, we confine ourselves to the WNT, which has been used extensively in the study of granular matter. For a Brownian particle, the steady-state solution of the Fokker-Planck equation is always a Gaussian velocity distribution, provided the noise is independent at different times, i.e., it does not depend upon the noise distribution. Therefore, any deviation from the Gaussian distribution (as demonstrated later) is not an artefact of the choice of thermostat.

We apply the algorithm suggested in Ref.~\cite{wm96}. During the simulations, we heat the system after a time step $dt$ by adding to the velocity of the $i^{\rm th}$ particle a random increment, which corresponds to the heating by noise,
\begin{eqnarray}
 \label{velupd}
 v_{i,\alpha}(t+dt) = v_{i,\alpha}(t) + \sqrt{r}\sqrt{dt}\varphi_{i,\alpha},
\end{eqnarray}
where $\alpha = x, y, z$. The random number $\varphi$ is uniformly distributed within the interval $[-0.5, 0.5]$ and $r$ is the amplitude of the noise, $r = 12\xi^2$. After the change in velocities, we set the center of mass velocity to zero, i.e.,
\begin{eqnarray}
 \label{veleq}
 \vec v_i \rightarrow \vec v_i - \frac{1}{N}\sum_{i = 1}^N \vec v_i.
\end{eqnarray}

It is useful to compare our model, in the sense of energy loss due to dissipation, with the commonly-used inelastic hard sphere (IHS) model for granular matter. In the IHS model, we consider a set of $N$ identical hard spheres with velocities $\vec{v}_i~(i = 1 \rightarrow N)$. These spheres undergo pair-wise inelastic collisions with the following rule:
\begin{eqnarray}
\label{ihs}
{\vec{v}_i}^{~\prime} &=& \vec{v}_i - \frac{1+e}{2} \left[ \hat{n} \cdot (\vec{v}_i - \vec{v}_j) \right] \hat{n} , \nonumber \\
{\vec{v}_j}^{~\prime} &=& \vec{v}_j + \frac{1+e}{2} \left[ \hat{n} \cdot (\vec{v}_i - \vec{v}_j) \right] \hat{n} .
\end{eqnarray}
Here, the primed variables denote the velocities of particles $i$ and $j$ after collision; and $\hat{n}$ is the unit vector pointing from $j$ to $i$ at the time of collision. The important physical parameter is the restitution coefficient $e$. The elastic limit corresponds to $e=1$, and $0<e<1$ for granular matter. At early times, when the system is homogeneous, the average kinetic energy or the temperature $T(t)$ cools according to Haff's law \cite{ph83}:
\begin{equation}
\label{haff}
T(t) = T_0 \left[1+ \frac{\epsilon}{2d} \omega (T_0) t\right]^{-2} .
\end{equation}
Here, $\epsilon = 1-e^2$, $d$ is the dimensionality, and $\omega (T_0)$ is the collision frequency at the initial temperature $T_0$. This has the approximate form \cite{cc70}
\begin{equation}
\omega (T) = \pi^{-1/2} \Omega_d \chi (n) n \sigma^{d-1} T^{1/2} ,
\end{equation}
where $\Omega_d = 2\pi^{d/2}/\Gamma(d/2)$ is the overall solid angle, and $\chi(n)$ is the pair correlation function at contact for hard spheres of number density $n$ and diameter $\sigma$. Clearly, the IHS model is not well-defined for dense granular matter, where many-body collisions play an important role.

In our model, we consider energy dissipation due to frictional forces. In dense granular media, the system evolves via rubbing of particles against each other. In this context, it is more natural to study dissipation due to Coulombic friction. For a gas of frictional hard spheres, we can derive the counterpart of Haff's law as follows. Consider a collision or rubbing of two grains, with fixed frictional force $\mu N$ and contact distance $b$. The loss of kinetic energy is the work done by the frictional force:
\begin{equation}
\Delta E = -\mu N b .
\end{equation}
The granular temperature is defined as $T = \langle v^2 \rangle/d$, where $\langle v^2 \rangle$ is the mean-squared velocity. Thus
\begin{equation}
\label{frhs}
\frac{dT(t)}{dt} = -\frac{\mu N b}{d} \omega (T) = -\frac{\mu N b}{d} \omega (T_0) \sqrt{\frac{T}{T_0}} .
\end{equation}
This equation is integrated to obtain
\begin{equation}
\label{frcool}
T(t) = T_0 \left[1-\frac{\mu N b}{2d T_0} \omega (T_0) t\right]^2 .
\end{equation}
At early times, the cooling law in Eq.~(\ref{frcool}) is analogous to Haff's law with $\mu \propto 1-e^2$. Clearly, as with Haff's law, this equation is only valid at early times, when the system is in the {\it homogeneous cooling state} (HCS) \cite{dps16}.

For a soft potential $V(r)$, as considered by us in Eq.~(\ref{soft}), the collision process is more complicated. The distance of closest approach $r_c$ is determined by $V(r_c) = T$, as $T$ measures the contribution to kinetic energy by the radial component of the collision velocity. The particles remain in contact over a distance $b = 2\sqrt{R_2^2 - r_c^2}$. The work done by friction during the collision is estimated by the integral $\int_{R_2}^{r_c} \mu V'(r) dr = \mu V(r_c) = \mu T$. Replacing this in Eq.~(\ref{frhs}), we obtain
\begin{equation}
\label{frss}
\frac{dT(t)}{dt} = -\frac{\mu}{d} \omega (T_0) \frac{T^{3/2}}{T_0^{1/2}} .
\end{equation}
This can be integrated to obtain the usual form of Haff's law with $\epsilon \rightarrow \mu$:
\begin{equation}
\label{frhaff}
T(t) = T_0 \left[1+ \frac{\mu}{2d} \omega (T_0) t\right]^{-2} .
\end{equation}
These cooling laws will prove useful in interpreting the numerical results presented later.

\subsection{Characterization of the Velocity Distribution Function}\label{sec2b}

In this subsection, we introduce the tools, which will enable us to characterize the VDF as it develops from its initial condition and settle into the steady state. We define an effective temperature from the mean kinetic energy of particles~\cite{bp04,pl01}. If the system is in a homogeneous state, i.e., density is uniform and the velocities are random, the time-dependent VDF is position independent and the time-dependent granular temperature, $T(t)$ can be written as
\begin{eqnarray}
 \label{temp}
 T(t) = \frac{2}{3n}\int d\vec v~\frac{mv^2}{2}f(\vec v, t),
\end{eqnarray}
where $n$ is the number density of particles~\cite{bp04}. We rescale the velocity of each particle by the thermal velocity
\begin{eqnarray}
 \label{tv}
 v_{\rm T}(t) = \sqrt{\frac{2T(t)}{m}},
\end{eqnarray}
and the distribution function is scaled as 
\begin{eqnarray}
 \label{ndist}
 f(\vec v, t) = \frac{n}{v_{\rm T}(t)^3}\tilde f(c),
\end{eqnarray}
where $c = |~\vec c~|$ is the normalized velocity defined as $\vec c=\vec v/v_{\rm T}(t)$.

We expect that starting from any initial condition of the nature described above and in the absence of friction and noise, the system will settle into a Maxwell-Boltzmann (MB) distribution function $f_\text{MB}(c)$,
\begin{eqnarray}
 \label{pmb}
 f_\text{MB}(c) = \frac{1}{\pi^{3/2}}\exp(-c^2).
\end{eqnarray}
We will start our system with the above distribution and evolve it in the presence of friction and noise. It would have been tempting to think that the only change in the VDF as it evolves in time is the change of $v_{\rm T}(t)$, so that the scaled distribution, Eq.~(\ref{pmb}), remains unchanged. This is not the case and even the eventual steady state is not a Gaussian. Thus, we allow for deviations from the initial scaled distribution by expanding $\tilde f(c)$ into a series of orthogonal functions as 
\begin{eqnarray}
 \label{exp}
 \tilde f(c) = f_\text{MB}(c)\sum_{p=0}^{\infty} a_pS_p(c^2),
\end{eqnarray}
where $a_p$'s are time-dependent Sonine coefficients with $a_0=1$ and $S_p(c^2)$ is the Sonine polynomial function~\cite{bp04}. The Sonine polynomials are defined as the associated Laguerre polynomial
\begin{eqnarray}
\label{laguerre}
S_p^m(c^2) = \sum_{n=0}^{p}\frac{\left(-1\right)^n\left(m + p\right)!}{n!\left(p-n\right)!\left(m + n\right)!}c^{2n}, \quad m = \frac{d}{2}-1 .
\end{eqnarray}

Notice that the expansion in Eq.~(\ref{exp}) assumes a Maxwellian behavior in the tail, as the $\exp(-c^2)$ factor is dominant at large $c$. This is found to be in good agreement with the numerical results we shall present shortly. In this context, it is relevant to point out that there have been several earlier studies of the VDF of a single particle under the influence of Coulombic friction and noise, starting with the work of de Gennes \cite{dg05} and Hayakawa \cite{hh05}. We have recently solved this problem for the case of arbitrary $d$ \cite{dps17}. In these studies, the VDF shows an exponential tail, which has been confirmed in the experiments of Gnoli {\it et al.} \cite{gph13}. Our present problem is somewhat more complicated as we consider the many-body case. Further, the particles interact via a conservative potential as well as dissipative Coulomb friction. Therefore, we expect a mixed distribution of the MB and exponential types. Our numerical results presented later show only a small deviation from the MB function in the tail regime.

In $d=3$, the index $m = 1/2$ and for the sake of simplicity, we write $S_p^{1/2}(c^2)=S_p(c^2)$. Therefore,
\begin{eqnarray}
 \label{sonine}
 S_p(c^2) = \sum_{n=0}^{p}\frac{\left(-1\right)^n\left(\frac{1}{2} + p\right)!}{n!\left(p-n\right)!\left(\frac{1}{2} + n\right)!}c^{2n}
\end{eqnarray}
with the orthonormality condition:
\begin{eqnarray}
 \label{ortho}
 \int d\vec c ~f_\text{MB}(c)S_p(c^2)S_q(c^2) = \frac{2\left(p + \frac{1}{2}\right)!}{\sqrt{\pi}~p!} = N_p\delta_{pq},
\end{eqnarray}
and the $p^{\rm th}$ Sonine coefficient is given by
\begin{eqnarray}
 \label{scoef}
 a_p = \frac{1}{N_p}\int d\vec c ~S_p(c^2)\tilde f(c).
\end{eqnarray}
The Sonine coefficients can be expressed in terms of the moments of $c$. The first few Sonine coefficients are given by
\begin{align}
 \label{coef1}
 a_1(t) &= 1 - \frac{\langle c^2\rangle}{\langle c^2\rangle_\text{MB}} = 0,\\
 \label{coef2}
 a_2(t) &= -1 + \frac{\langle c^4\rangle}{\langle c^4\rangle_\text{MB}},\\ 
 \label{coef3}
 a_3(t) &= 1 + 3a_2 - \frac{\langle c^6\rangle}{\langle c^6\rangle_\text{MB}},\\
 \label{coef4}
 a_4(t) &= -1 - 6a_2 + 4a_3 + \frac{\langle c^8\rangle}{\langle c^8\rangle_\text{MB}},\text{etc.}
\end{align}
Here, we define $\langle c^{2k}\rangle$ as
\begin{eqnarray}
 \label{Nmom}
 \langle c^{2k}\rangle = \int d\vec c ~c^{2k}\tilde f(c),
\end{eqnarray}
For the MB distribution function, this quantity is given by
\begin{eqnarray}
 \label{MBmom}
 \langle c^{2k}\rangle_\text{MB} = \frac{\Gamma\left(k + \frac{3}{2} \right)}{\Gamma\left(\frac{3}{2} \right)}.
\end{eqnarray}
All $a_i$'s are zero for the MB distribution for $i>1$. The nonzero Sonine coefficients $a_2$, $a_3$ and so on measure the deviation from the MB distribution. 

The system is initialized by randomly placing particles in the simulation box, such that there is no overlap between the cores of any two particles. At $t=0$, all these particles have the same speed but the velocity vector points in random directions so that $\sum_{i=1}^N \vec v_i = 0$. The system is allowed to evolve till $t=50$ with $\mu=0$ and $\xi=0$ i.e., the elastic limit without noise. The system is relaxed to a MB velocity distribution, which serves as the initial condition for our simulation of frictional spheres with $\mu \ne 0$ and with noise,  $\xi\ne0$. We use $\mu=0.10$ for all our simulations, and all the results presented here are obtained as an average over ten independent runs. We consider systems with volume fractions $\phi=0.10$ ($L=122.4$) and $\phi=0.40$ ($L=77$), below the jamming volume fraction $\phi_{\rm J}\simeq 0.47$ in our model.

\section{Numerical Results}\label{sec3}

\begin{figure}
\centering
\includegraphics*[width=0.90\textwidth]{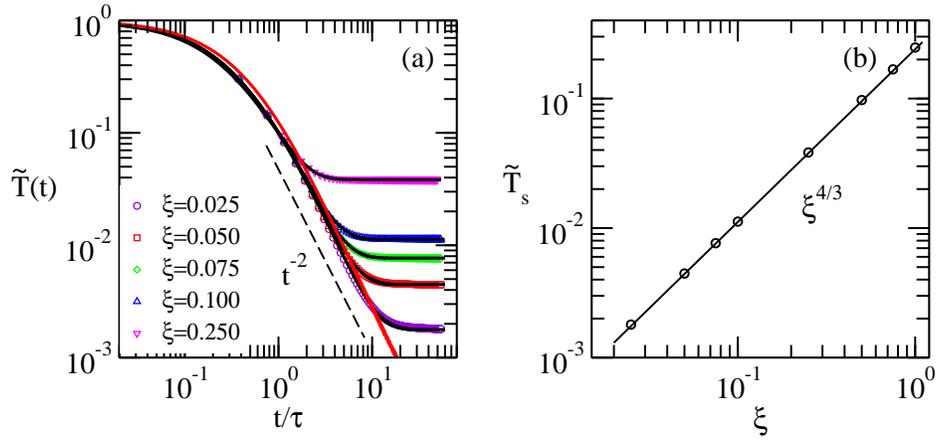}
\caption{\label{fig1} (a) Time evolution of scaled granular temperature $\tilde T(t)=T(t)/T(0)$ for small volume fraction $\phi=0.10$. Plot of $\tilde T(t)$ vs. $t/\tau$ on a $\log$-$\log$ scale for different driving force amplitude $\xi$ as mentioned. Numerical data is shown by points. The solid red line represents the decay of $\tilde T(t)$ for the free evolution, i.e., when $\xi=0$. Solid black lines on different data sets represent the solution of Eq.~(\ref{ncol}) for the corresponding values of $\xi$. The dashed line marked with $t^{-2}$ represents the analytically obtained cooling law. (b) Plot of steady state scaled granular temperature $\tilde T_s$ vs. $\xi$ on a $\log$-$\log$ scale. The solid line marked with $\xi^{4/3}$ represents the analytically obtained relation between $\tilde T_s$ and $\xi$.}
\end{figure}

At $t=0$, we start with homogeneous density field and random velocity field in thermal equilibrium. Particles in the system lose their kinetic energy due to the frictional interaction, while gaining energy from the noise. In Fig.~\ref{fig1}(a), we show the time evolution of the scaled temperature $\tilde T(t)[=T(t)/T(0)]$ vs. $t/\tau$ for $\phi=0.10$ and different values $\xi$. Details are given in the caption of the figure. Here, $\tau$ is the time-scale at which $\tilde T(t)$ falls to $0.1$. At early times, the system cools due to frictional interaction among the particles. For low-density gases with $\phi=0.10$, the early time cooling is similar for freely evolving and heated cases. We observe a power-law decay as $\tilde T(t)\sim t^{-\alpha}$ with approximate decay exponent $\alpha=2$. Analytically, this can be obtained as follows. In the low-density limit, the particle-particle interaction is small. Assuming the system remains homogeneous at early stage of evolution, $T(t)$ decays as
\begin{eqnarray}
  \label{cool}
  \frac{dT}{dt}=-kT^{3/2},
\end{eqnarray}
where $k = (\mu/d) (\omega(T_0)/\sqrt{T_0})$ from Eq.~(\ref{frss}). Recently, Burton \textit{et al.}~\cite{bln13} experimentally studied the cooling properties of colliding solid CO$_2$ particles. They observed that the kinetic energy which manifests the granular temperature decays in a power-law fashion as $E(t)\sim t^{-\gamma}$ with $\gamma=1.8\pm 0.2$. Thus our numerically-obtained decay exponent is consistent with their experimental data. In the presence of noise, Eq.~(\ref{cool}) becomes
\begin{eqnarray}
  \label{ncol}
  \frac{dT}{dt}=m\xi^2-kT^{3/2}.
\end{eqnarray}
Here, $m\xi^2$ corresponds to the rate of increase of temperature due to noise. We numerically solve Eq.~(\ref{ncol}); and solid black lines in Fig.~\ref{fig1}(a) denote the solution for different values of $\xi$. These are in good agreement with the numerical data. In the steady state $dT/dt=0$, which gives $T_{\rm s}\sim \xi^{4/3}$. In Fig.~\ref{fig1}(b), we show the variation of scaled steady state granular temperature $\tilde T_{\rm s}$ as a function of the noise amplitude $\xi$. We find that $\tilde T_{\rm s}$ follows a power-law as $\tilde T_{\rm s}\sim\xi^{4/3}$. Therefore, our numerical data in Fig.~\ref{fig1}(b) is in good agreement with the analytical prediction. Also, for any initial temperature $T_0$, $dT/dt$ will be positive for the driving force amplitude $\xi>\xi_0=\sqrt{k/m}~T_0^{3/4}$. Therefore, for $\xi>\xi_0$ the system will heat up and reach a steady state with $T_{\rm s}>T_0$.

\begin{figure}
\centering
\includegraphics*[width=0.90\textwidth]{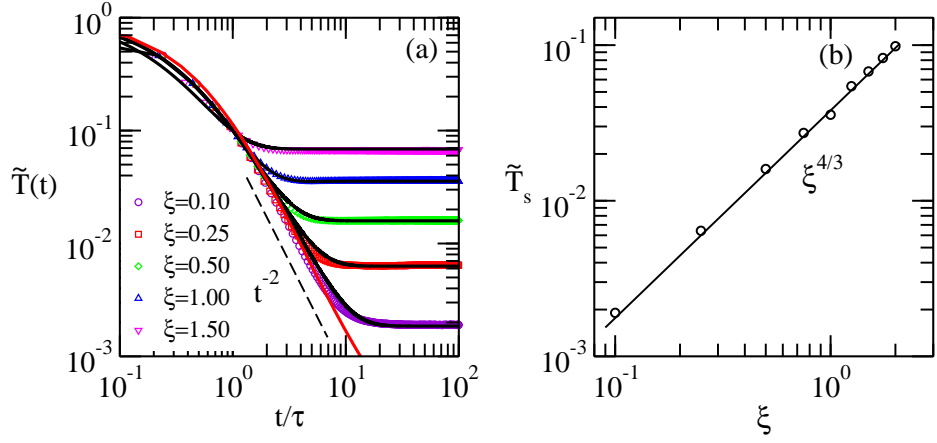}
\caption{\label{fig2} Analogous to Fig.~\ref{fig1} for dense granular fluid with $\phi=0.40$. (a) Plot of $\tilde T(t)$ vs. $t/\tau$ on a $\log$-$\log$ scale for different $\xi$ as mentioned. Numerical data is shown by points. The solid red line represents the decay of $\tilde T(t)$ for $\xi=0$. Solid black lines on different data sets represent the solution of Eq.~(\ref{ncol}) for the corresponding values of $\xi$. The dashed line represents the $t^{-2}$ cooling law. (b) Plot of $\tilde T_s$ vs. $\xi$ on a $\log$-$\log$ scale. The solid line marked with $\xi^{4/3}$ represents the analytical result.}
\end{figure}

For dense granular fluids with $\phi=0.40$, the early stage cooling is also described by $\tilde T(t)\sim t^{-2}$, as shown in Fig.~\ref{fig2}(a) for all values of $\xi$. As in the low-density case, the early stage cooling in heated fluids is similar to the freely cooling fluid. At the late stage, $\tilde T(t)$ reaches a steady state value $\tilde T_{\rm s}$ by balancing the dissipation of energy due to frictional interactions and the input energy due to heating. As before, the solution of Eq.~(\ref{ncol}) and the numerical data agree well for a given value of $\xi$. In Fig.~\ref{fig2}(b), we plot the dependence of $\tilde T_{\rm s}$ with $\xi$. Similar to the low-density case as described in Fig.~\ref{fig1}(b), we find that $\tilde T_{\rm s}$ follows the identical power-law: $\tilde T_{\rm s}\sim\xi^{4/3}$. From Fig.~\ref{fig1}(b) and Fig.~\ref{fig2}(b), it is clear that for a given $\xi$, $\tilde T_{\rm s}$ depends on the density $\phi$. $\tilde T_{\rm s}$ is lower for higher $\phi$ as dissipation is more effective in the denser system due to large number of particles rubbing against each particle.

\begin{figure}
\centering
\includegraphics*[width=0.50\textwidth]{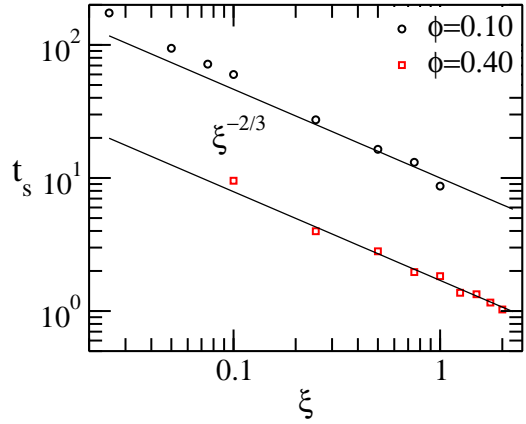}
\caption{\label{fig3} Plot of $t_s$ vs. $\xi$ on a $\log$-$\log$ scale for different values of $\phi$, as denoted by the symbols indicated. The solid lines marked with $\xi^{-2/3}$ represent the analytically obtained power-law decay of $t_s$, irrespective of $\phi$.}
\end{figure}

Next, we calculate the crossover time, $t_{\rm s}$ at which system reaches steady state. To obtain $t_{\rm s}$ from the numerical data, we calculate the instantaneous exponent $\beta$ for $\tilde T(t)$ vs. $t$. This is computed as $\beta= -d\ln\tilde T(t)/d\ln t$, and we define $t_{\rm s}$ as the time at which $\beta$ becomes $0.01$. In Fig.~\ref{fig3}, we plot $t_{\rm s}$ vs. $\xi$ for different values of $\phi$. Details are given in the figure caption. Clearly, for a given value of $\xi$, $t_{\rm s}$ is smaller for denser systems due to large number of interacting partners. We find that $t_{\rm s}$ decays as $t_{\rm s}\sim\xi^{-\theta}$ with $\theta\sim 0.66$. Analytically, we have 
\begin{eqnarray}
\label{steady}
T(t_{\rm s})\sim t_{\rm s}^{-2}\sim \xi^{4/3}.
\end{eqnarray}
This gives $t_{\rm s}\sim \xi^{-2/3}$. Thus our numerically estimated decay exponent $\theta$ is approximately equal to the analytical value.

\begin{figure}
\centering
\includegraphics*[width=0.80\textwidth]{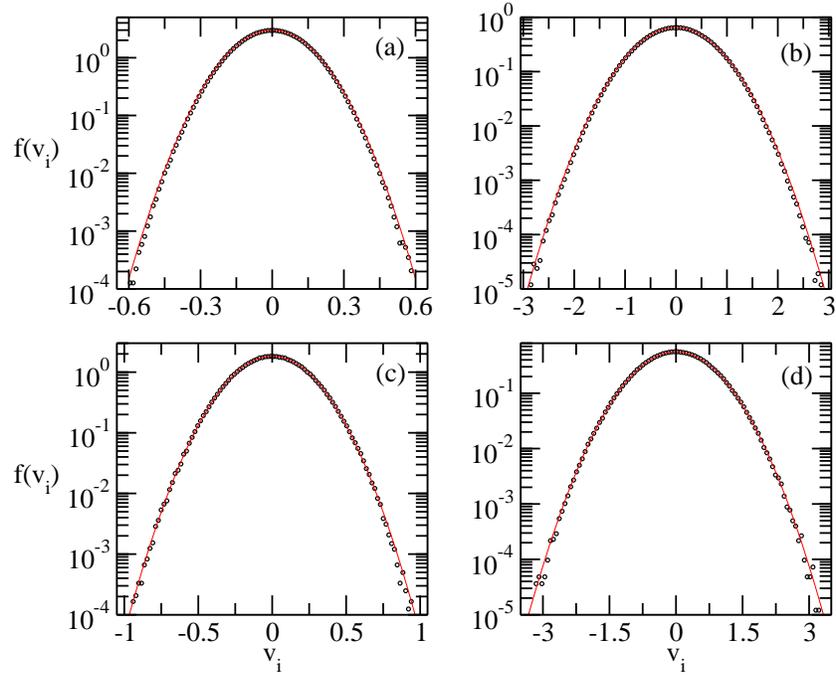}
\caption{\label{fig4} Plot of the steady state velocity distribution functions $f(v_i)$ for different values of $\phi$ and $\xi$: (a) $\phi=0.10$ and $\xi=0.025$, (b) $\phi=0.10$ and $\xi=0.25$, (c) $\phi=0.40$ and $\xi=0.25$, (d) $\phi=0.40$ and $\xi=1.50$. Open circles represent numerical data obtained by averaging over ten independent runs and over steady state distributions. The solid line in each figure represents the scaled MB distribution for the corresponding steady state temperature.}
\end{figure}

Next, we study the steady state VDF. The natural framework to study the velocity distributions for the unheated elastic granular gas ($\xi=0$, $\mu=0$) is the Boltzmann equation. In this case, any arbitrary initial velocity distribution rapidly evolves (after certain time) to the MB distribution:
\begin{eqnarray}
 \label{mbd}
 P_{\rm MB}(\vec v) = \left(\frac{1}{\pi v_0^2}\right)^{3/2}\exp\left(-\frac{\vec v^{\,2}}{v_0^2}\right),~~~~~~v_0^2=\frac{2\langle \vec v^{\,2}\rangle}{3}.
\end{eqnarray}
where $\vec v=(v_x,v_y,v_z)$ is the velocity of the particle. For heated frictional granular gases with $\xi \ne 0$ and $\mu \ne 0$, the VDF is different from the MB distribution given by Eq.~(\ref{mbd}). Fig.~\ref{fig4} shows the VDFs for different values of $\phi$ and $\xi$ in the steady state. Numerical details are given in the figure caption. Clearly, the numerically-obtained data deviates slightly from the MB distribution at the steady-state temperature, primarily in the tail region. However, the tail behavior is not of the exponential form, $\ln f(v_i) \sim -v_i$, as is clear from the linear-logarithmic plot in Fig.~\ref{fig4}. We characterize the deviation by calculating the Sonine polynomial coefficients given by Eqs.~(\ref{coef2})-(\ref{coef4}). 

\begin{figure}
\centering
\includegraphics*[width=0.85\textwidth]{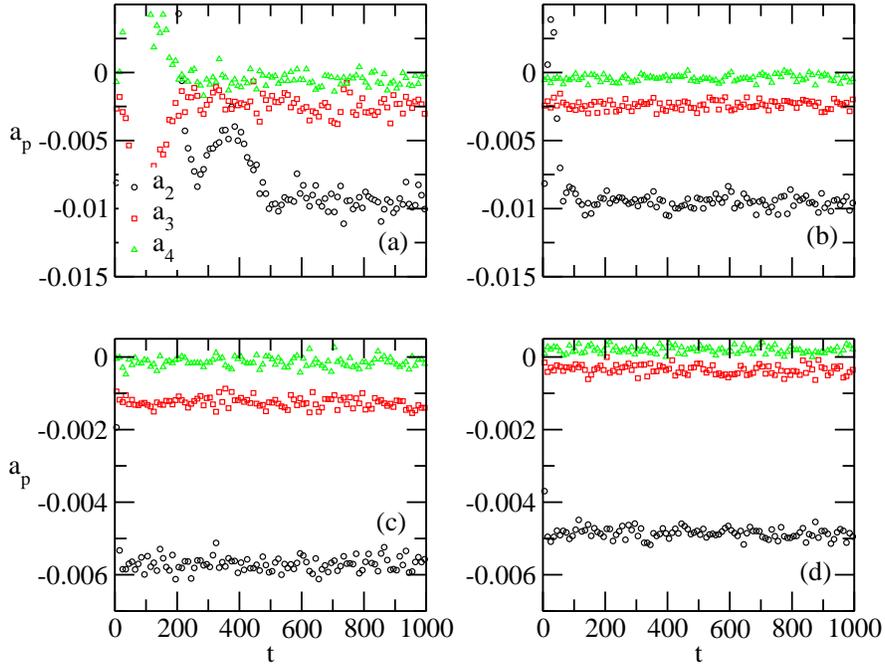}
\caption{\label{fig5} Time evolution of Sonine coefficients, $a_2$, $a_3$ and $a_4$ for different values of $\phi$ and $\xi$. Numerical data is obtained by averaging over ten independent runs. After early transient, $a_p$'s settle into a constant. (a) $\phi=0.10$ and $\xi=0.025$, (b) $\phi=0.10$ and $\xi=0.25$, (c) $\phi=0.40$ and $\xi=0.25$, (d) $\phi=0.40$ and $\xi=1.50$.}
\end{figure}

In Fig.~\ref{fig5}, we plot $a_2$, $a_3$, and $a_4$ for different $\phi$ and $\xi$. Simulation details are given in the figure caption. Clearly, in the steady state, $a_p$'s settle into a constant value. Also, the amplitude of $a_3$ is one order of magnitude smaller than $a_2$ and $a_4$ is one order of magnitude smaller than $a_3$ and so on, suggesting the convergence of the Sonine polynomial expansion. The constant values of Sonine coefficients imply that we have reached steady state and the form of the VDF is different from the MB distribution for all cases. In the low-density limit shown in (a)-(b), the evolution is governed by two-body collisions. The noise intensity only sets the temperature scale and the Sonine coefficients are independent of the noise strength, as expected. For higher densities shown in (c)-(d), we expect multi-particle collisions to play a significant role. In this case, the Sonine coefficients show a weak dependence on the noise strength.

As mentioned earlier, Eqs.~(\ref{haff}) and (\ref{frhaff}) suggest the analogy $\mu \simeq \epsilon = 1-e^2$ between the frictional gas and the inelastic hard-sphere gas. For the latter, there exist accurate analytical predictions for $a_2, a_3$ as a function of $e$ \cite{ms00,ne98,gm2009}. In the low-density limit appropriate to the granular gas, we find that $a_2$ and $a_3$ for $\mu = 0.10$ [shown in Figs.~\ref{fig5}(a)-(b)] are comparable to the analytical values \cite{gm2009} for $e=0.95 \simeq \sqrt{1-\mu}$. This demonstrates a strong analogy between the frictional gas studied here and the hard-sphere gas. A more precise derivation of Sonine coefficients for a frictional gas is a demanding project, and we defer it for subsequent work.

\begin{figure}
\centering
\includegraphics*[width=0.90\textwidth]{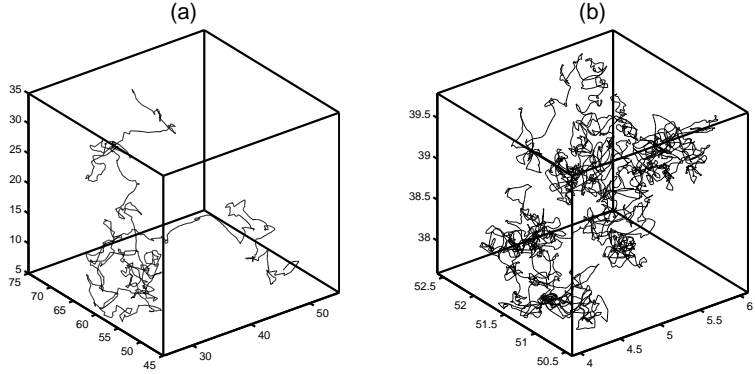}
\caption{\label{fig6} Plot of single particle trajectories up to $t=500$ for $\xi=0.25$ and (a) $\phi=0.10$ and (b) $\phi=0.40$ respectively. Clearly, the spatial volume covered by the motion of the particle for high-density system is much smaller than that compared to the low-density case.}
\end{figure}

Next, we consider the diffusive motion of tagged particles. In Fig.~\ref{fig6}(a), we plot the trajectory of a tagged particle for $\phi=0.10$ and $\xi=0.25$. In this case, the particle performs a random walk in the simulation box. However, for higher volume fractions with $\phi=0.40$ and the same $\xi$, the tagged particle is trapped in the temporary cages formed by its neighbors for small time and makes random jumps from one cage to another causing local rearrangement of particles as shown in Fig.~\ref{fig6}(b). This may be interpreted as a {\it Levy flight}, which has a distribution of step lengths. The largest of these lengths corresponds to the jumps between local cages, which have been studied in the context of glassy dynamics \cite{bk07,gk09,fhv16}. The continuous interaction of the tagged particle with its neighbors gives rise to the zig-zag trajectory. Clearly, the spatial volume covered by the particle for the dilute limit is higher than the dense limit. Reis {\it et al.}~\cite{ris07} experimentally observed the similar dynamics in driven two-dimensional granular fluids.

\begin{figure}
\centering
\includegraphics*[width=0.95\textwidth]{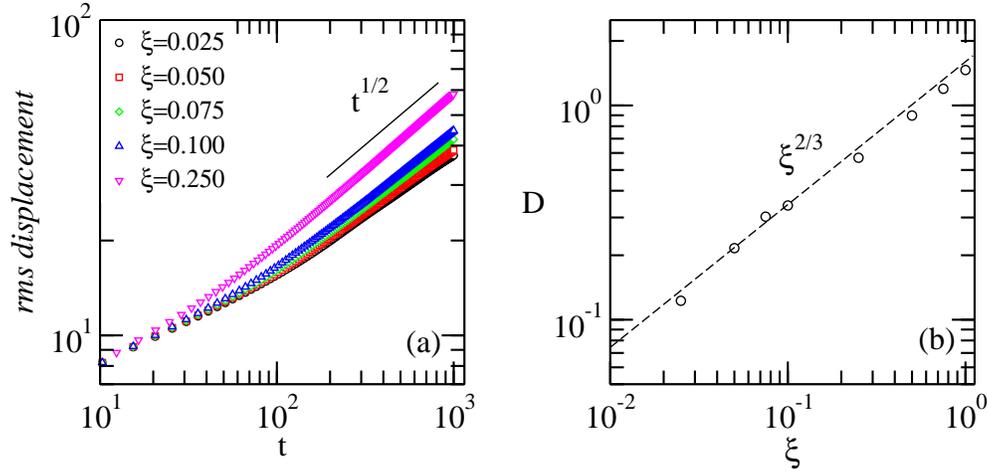}
\caption{\label{fig7} Plot of rms displacement and self-diffusion coefficient for $\phi=0.10$. (a) $rms~displacement$ vs. $t$ on a $\log$-$\log$ scale for different values of $\xi$. After early transient, the root mean-squared displacement of the particles increases as $t^{1/2}$. The solid line marked with $t^{1/2}$ represents the diffusive motion. (b) Plot of $D$ vs. $\xi$ on a $\log$-$\log$ scale. The dashed line marked with $\xi^{2/3}$ shows $D$ increases with $\xi$ as $D\sim\xi^{2/3}$.}
\end{figure}

Next, we characterize the particle dynamics in these heated granular systems by calculating the root mean-squared (rms) displacement, $\sqrt{\left\langle r^2(t)\right\rangle}$ and the average self-diffusion coefficient, $D$ for different values of $\phi$ and $\xi$. In Fig.~\ref{fig7}(a), we plot {\it rms displacement} vs. $t$ for $\phi=0.10$. Numerical details are described in the figure caption. In this case sufficient phase space is available and the particles perform uncorrelated motion which gives rise to the diffusive motion as $\left\langle r^2(t)\right\rangle=6Dt$. We calculate $D$ numerically using the relation
\begin{eqnarray}
	\label{diff}
	D=\frac{1}{6}\frac{d}{dt}\left\langle r^2(t)\right\rangle.
\end{eqnarray}
In Fig.~\ref{fig7}(b), we plot $D$ vs. $\xi$ for the same volume fraction $\phi=0.10$. We observe that $D$ increases with $\xi$ as $D\sim\xi^{2/3}$. Kawarada and Hayakawa~\cite{kh04} found that the particles show diffusive motion in the presence of solid friction and noise with $D\sim\sqrt{T}$. Here in the steady state, we expect $D\sim\sqrt{T_s}$, i.e., $D\sim\xi^{2/3}$. Thus, our numerical data is in good agreement with the analytical prediction.

\begin{figure}
\centering
\includegraphics*[width=0.95\textwidth]{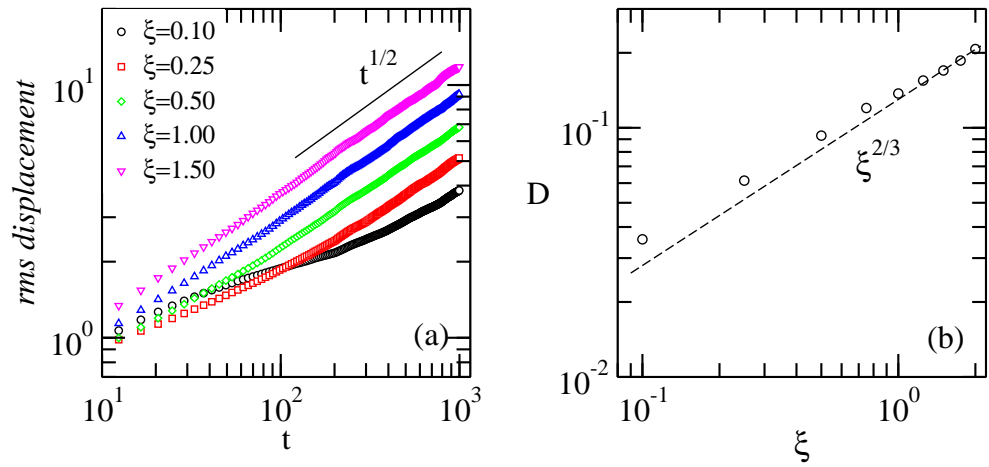}
\caption{\label{fig8} (a) Log-$\log$ plot of $rms~displacement$ as a function of time, $t$ for different $\xi$ as mentioned. The solid line labeled with $t^{1/2}$ indicates the diffusive motion of particles. (b) Log-$\log$ plot of $D$ as a function of $\xi$. The dashed line labeled with $\xi^{2/3}$ represents analytically obtained relation.}
\end{figure}

For a dense granular fluid with $\phi=0.40$, we plot {\it rms displacement} vs. $t$ in Fig.~\ref{fig8}(a) for different values of $\xi$. Recall that this case corresponds to the Levy flight depicted in Fig.~\ref{fig6}(b). At early times, particles are stuck in the temporary cages formed by its neighbors. At the late stage of evolution, the particles make random jumps from one cage to the another by making local rearrangement of the neighboring particles. These uncorrelated jumps give rise to the diffusive motion on the average with $\left\langle r^2(t)\right\rangle\sim t$ for all values of $\xi$. However, the scale of motion is much smaller than in Fig.~\ref{fig7}(a). Notice that the early-time behavior for the weakest noise-strength ($\xi = 0.10$) in Fig.~\ref{fig8}(a) explicitly shows slowing-down due to trapping. Figure~\ref{fig8}(b) shows variation of $D$ with $\xi$ for $\phi=0.40$. Similar to the low-density case, we found the power-law $D\sim\xi^{2/3}$. However, $D$ is one order of magnitude smaller in the high-density case because of the lack of available phase space for particle motion. For $\phi>\phi_{\rm J}$, rms displacement becomes constant which implies particles become trapped in the permanent cages formed by their neighbors.

\section{Conclusion}\label{sec4}

Let us conclude this paper with a summary and discussion of our results. We have studied the dynamical properties of heated granular fluids using large-scale molecular dynamics simulation in three dimensions. We consider the solid friction between any pair of interacting particles as the only dissipation mechanism of energy. The system is heated by using white-noise thermostat. At the early stage of evolution, the system cools down with time, and the cooling is similar to the free evolution. Namely, the effective temperature decays in time as $\tilde T(t)\sim t^{-2}$, irrespective of volume fraction $\phi$. However, at the late stage, the system attains a steady state temperature $T_s$ by balancing the input energy due to heating and dissipation due to friction. We found a power-law dependence between $T_s$ and $\xi$ as $T_s\sim\xi^{4/3}$, valid for moderate values of $\phi$. Also the time $t_s$ at which system reaches steady state shows a power-law dependence with $\xi$ as $t_s\sim\xi^{-2/3}$, irrespective of $\phi$.

In the steady state, velocity distribution shows deviation from the Maxwell-Boltzmann (MB) distribution, which has been studied by using a Sonine polynomial expansion of the velocity distribution. The nonzero value of most significant Sonine coefficient, $a_2$, measures the deviation from MB distribution is few order of magnitude higher than $a_3$, $a_4$ and so on. The decreasing order of higher order Sonine coefficients confirms the convergence of Sonine polynomial expansion. Single particle dynamics shows that particles follow diffusive motion in the low-density limit. However, in the high-density limit, we observe caging dynamics of a given particle which is arrested by its neighbors and random jumps of particles from one cage to another gives diffusive motion at later times. The self-diffusion coefficient $D$ increases with $\xi$ as $D\sim\xi^{2/3}$, for all $\phi<\phi_{\rm J}$ and it is consistent with the analytical prediction.

\subsubsection*{Acknowledgments}

PD acknowledges financial support from Council of Scientific and Industrial Research, India. The research of MS, grant number 839/14, was supported by the ISF within the ISF-UGC joint research program framework.

\subsubsection*{Conflict of Interest}

The authors declare that they have no conflict of interest.

\newpage

\newpage
\begin{thebibliography}{99}

\bibitem{jn96} H.M. Jaeger and S.R. Nagel, Granular solids, liquids, and gases, Rev. Mod. Phys. \textbf{68}, 1259-1273 (1996).

\bibitem{dg99} P.G. de Gennes, Granular matter: a tentative view, Rev. Mod. Phys. \textbf{71}, S374-S382 (1999).
 
\bibitem{at06} I.S. Aranson and L.S. Tsimring, Patterns and collective behavior in granular media: Theoretical concepts, Rev. Mod. Phys. \textbf{78}, 641-692 (2006).
 
\bibitem{jd94} J. Duran, \textit{Sands, Powders and Grains: An Introduction to the Physics of Granular Materials}, Springer-Verlag, New York (1994).

\bibitem{bp04} N.V. Brilliantov and T. Poschel, \textit{Kinetic Theory of Granular Gases}, Oxford University Press, Oxford (2004).
 
\bibitem{dps16} P. Das, S. Puri and M. Schwartz, Clustering and velocity distributions in granular gases cooling by solid friction, Phys. Rev. E \textbf{94}, 032907 (2016).
  
\bibitem{bes10} R. Blumenfeld, S.F. Edwards and M. Schwartz, da Vinci fluids, catch-up dynamics and dense granular flow, Euro. Phys. J. E \textbf{32}, 333-338 (2010).
 
\bibitem{sb11} M. Schwartz and R. Blumenfeld,  Plug flow formation and growth in da Vinci Fluids, Granul. Matter \textbf{13}, 241-245 (2011).
 
\bibitem{pl01} \textit{Granular Gases}, edited by T. P\"{o}schel and S. Luding, Springer-Verlag, Heidelberg (2001).
 
\bibitem{dp03} S.K. Das and S. Puri, Kinetics of inhomogeneous cooling in granular fluids, Phys. Rev. E \textbf{68}, 011302 (2003).
 
\bibitem{sdsp03} S.K. Das and S. Puri, Pattern formation in the inhomogeneous cooling state of granular fluids, Euro. Phys. Lett. \textbf{61}, 749-755 (2003).
 
\bibitem{ap03} S.R. Ahmad and S. Puri, Velocity distributions in a freely evolving granular gas, Euro. Phys. Lett. \textbf{75}, 56-62 (2006).
 
\bibitem{ap07} S.R. Ahmad and S. Puri, Velocity distributions and aging in a cooling granular gas, Phys. Rev. E \textbf{75}, 031302 (2007).
 
\bibitem{ph83} P.K. Haff, Grain flow as a fluid-mechanical phenomenon, J. Fluid. Mech. \textbf{134}, 401-430 (1983).
 
\bibitem{gz93} I. Goldhirsch and G. Zanetti, Clustering instability in dissipative gases, Phys. Rev. Lett. \textbf{70}, 1619-1622 (1993).
 
\bibitem{gtz93} I. Goldhirsch, M.-L. Tan and G. Zanetti, A molecular dynamical study of granular fluids: the unforced granular gas, J. Sci. Comput. \textbf{8}, 1-40 (1993).
 
\bibitem{bkb15} N.V. Brilliantov, P.L. Krapivsky, A. Bodrova, F. Spahn, H. Hayakawa, V. Stadnichuk and J. Schmidt, Size distribution of particles in Saturn’s rings from aggregation and fragmentation, Proc. Natl. Acad. Sci. USA \textbf{112}, 9536-9541 (2015).
 
\bibitem{pgp03} S.C. du Pont, P. Gondret, B. Perrin and M. Rabaud, Granular avalanches in fluids, Phys. Rev. Lett. \textbf{90}, 044301 (2003).
 
\bibitem{gg14} N. Gravish and D.I. Goldman, Effect of volume fraction on granular avalanche dynamics, Phys. Rev. E \textbf{90}, 032202 (2014).
 
\bibitem{ms00} J.M. Montanero and A. Santos, Computer simulation of uniformly heated granular fluids, Granul. Matter \textbf{2}, 53-64 (2000).
 
\bibitem{ne98} T.P.C. van Noije and M.H. Ernst, Velocity distributions in homogeneous granular fluids: the free and the heated case, Granul. Matter \textbf{1}, 57-64 (1998).
 
\bibitem{sor09} J. Schmidt, K. Ohtsuki, N. Rappaport, H. Salo and F. Spahn, in \textit{Saturn from Cassini-Huygens, The
Structure of Saturn's Rings}, edited by M.K. Dougherty, L.W. Esposito and S. M. Krimigis, Springer, Heidelberg (2009), pp. 413-458.
 
\bibitem{bhl84} F.G. Bridges, A. Hatzes and D.N.C. Lin, Structure, stability and evolution of Saturn's rings, Nature \textbf{309}, 333-335 (1984).
 
\bibitem{gm04} G.D.R. Midi, On dense granular flows, Eur. Phys. J. E \textbf{14}, 341-365 (2004).
 
\bibitem{mus95} F. Melo, P.B. Umbanhowar and H.L. Swinney, Hexagons, kinks, and disorder in oscillated granular layers, Phys. Rev. Lett. \textbf{75}, 3838-3941 (1995).
 
\bibitem{ums96} P.B. Umbanhowar, F. Melo and H.L. Swinney, Localized excitations in a vertically vibrated granular layer, Nature \textbf{382}, 793-796 (1996).
 
\bibitem{gr00} G.H. Ristow, \textit{Pattern Formation in Granular Materials}, Springer, Heidelberg (2000).
 
\bibitem{zls94} O. Zik, D. Levine, S.G. Lipson, S. Shtrikman and J. Stavans, Rotationally induced segregation of granular materials, Phys. Rev. Lett. \textbf{73}, 644-647 (1994).
 
\bibitem{ph99} S. Puri and H. Hayakawa, Dynamical behaviour of rotated granular mixtures, Physica A \textbf{270}, 115-124 (1999); S. Puri and H. Hayakawa, Segregation of granular mixtures in a rotating drum, Physica A \textbf{290}, 218-242 (2001). 
 
\bibitem{bls01} L. Bocquet, W. Losert, D. Schalk, T. C. Lubensky and J.P. Gollub, Granular shear flow dynamics and forces: experiment and continuum theory, Phys. Rev. E \textbf{65}, 011307 (2001).
 
\bibitem{wp02} R.D. Wildman and D.J. Parker, Coexistence of two granular temperatures in binary vibrofluidized beds, Phys. Rev. Lett. \textbf{88}, 064301 (2002).
 
\bibitem{fm02} K. Feitosa and N. Menon, Breakdown of energy equipartition in a 2D binary vibrated granular gas, Phys. Rev. Lett. \textbf{88}, 198301 (2002).
 
\bibitem{ao02} I.S. Aranson and J.S. Olafsen, Velocity fluctuations in electrostatically driven granular media, Phys. Rev. E \textbf{66}, 061302 (2002).
 
\bibitem{sak05} A. Snezhko, I.S. Aranson and W.-K. Kwok, Structure Formation in electromagnetically driven granular media, Phys. Rev. Lett. \textbf{94}, 108002 (2005).
 
\bibitem{spe99} C. Salue$\tilde{n}$a, T. P\"{o}schel and S.E. Esipov, Dissipative properties of vibrated granular materials, Phys. Rev. E \textbf{59}, 4422-4425 (1999).
 
\bibitem{bt98} A. Barrat and E. Trizac, Lack of energy equipartition in homogeneous heated binary granular mixtures, Granul. Matter \textbf{4}, 57-63 (1998).
 
\bibitem{pmp02} R. Pagnani, U.M.B. Marconi and A. Puglisi, Driven low density granular mixtures, Phys. Rev. E \textbf{66}, 051304 (2002).

\bibitem{ms98} Y. Murayama and M. Sano, Transition from Gaussian to non-Gaussian velocity distribution functions in a vibrated granular bed, J. Phys. Soc. Jpn. \textbf{67}, 1826-1829 (1998).
 
\bibitem{po98} G. Peng and T. Ohta, Scaling and correlations in heated granular materials, J. Phys. Soc. Jpn. \textbf{67}, 2561-2564 (1998).
 
\bibitem{net99} T.P.C. van Noije, M.H. Ernst, E. Trizac and I. Pagonabarraga, Randomly driven granular fluids: large-scale structure, Phys. Rev. E \textbf{59}, 4326-4341 (1999).
 
\bibitem{kh04} A. Kawarada and H. Hayakawa, Non-Gaussian velocity distribution function in a vibrating granular bed, J. Phys. Soc. Jpn. \textbf{73}, 2037-2040 (2004).
 
\bibitem{wm96} D.R.M. Williams and F.C. MacKintosh, Driven granular media in one dimension: correlations and equation of state, Phys. Rev. E \textbf{54}, R9-R12 (1996); D.R.M. Williams, Driven granular media and dissipative gases: correlations and liquid-gas phase transitions, Physica A \textbf{233}, 718-729 (1996).

\bibitem{bdp12} A. Bodrova, A.K. Dubey, S. Puri and N.V. Brilliantov, Intermediate regimes in granular Brownian motion: superdiffusion and subdiffusion, Phys. Rev. Lett. {\bf 109}, 178001 (2012); A.K. Dubey, A. Bodrova, S. Puri and N.V. Brilliantov, Velocity distribution function and effective restitution coefficient for a granular gas of viscoelastic particles, Phys. Rev. E \textbf{87}, 062202 (2013).
 
\bibitem{at87} M.P. Allen and D.J. Tildesley, \textit{Computer Simulation of Liquids}, Oxford University Press, Oxford (1987).
 
\bibitem{fs02} D. Frenkel and B. Smit, \textit{Understanding Molecular Simulation: From Algorithms to Applications}, Academic Press, New York (2002).
 
\bibitem{dr04} D.C. Rapaport, \textit{The Art of Molecular Dynamics Simulation}, Cambridge University Press, Cambridge (2004).
 
\bibitem{as03} A. Santos, Transport coefficients of $d$-dimensional inelastic Maxwell models, Physica A {\bf 321}, 442-466 (2003).

\bibitem{hh03} H. Hayakawa, Hydrodynamics of driven granular gases, Phys. Rev. E {\bf 68}, 031304 (2003).

\bibitem{crg15} M.G. Chamorro, F. Vega Reyes and V. Garzo, Non-Newtonian hydrodynamics for a dilute granular suspension under uniform shear flow, Phys. Rev. E {\bf 92}, 052205 (2015).

\bibitem{cc70} S. Chapman and T.G. Cowling, {\it The Mathematical Theory of Non-uniform Gases}, Cambridge University Press, Cambridge (1970).
 
\bibitem{dg05} P.G. de Gennes, Brownian motion with dry friction, J. Stat. Phys. {\bf 119}, 953-962 (2005).
 
\bibitem{hh05} H. Hayakawa, Langevin equation with Coulomb friction, Physica D {\bf 205}, 48-56 (2005).
 
\bibitem{dps17} P. Das, S. Puri and M. Schwartz, Single particle Brownian motion with solid friction, Eur. Phys. J. E {\bf 40}, 60 (2017).

\bibitem{gph13} A. Gnoli, A. Puglisi and H. Touchette, Granular Brownian motion with dry friction, Euro. Phys. Lett. {\bf 102}, 14002 (2013);
A. Gnoli, A. Petri, F. Dalton, G. Pontuale, G. Gradenigo, A. Sarracino and A. Puglisi, Brownian ratchet in a thermal bath driven by Coulomb friction, Phys. Rev. Lett. {\bf 110}, 120601 (2013).
 
\bibitem{bln13} J.C. Burton, P.Y. Lu and S.R. Nagel, Collision dynamics of particle clusters in a two-dimensional granular gas, Phys. Rev. E \textbf{88}, 062204 (2013).

\bibitem{bk07} L. Berthier and W. Kob, The Monte Carlo dynamics of a binary Lennard-Jones glass-forming mixture, J. Phys. Cond. Matt. {\bf 19}, 205130 (2007).

\bibitem{gm2009} A. Santos and J. M. Montanero, The second and third Sonine coefficients of a freely cooling granular gas revisited, Granul. Matter {\bf 11}, 157-168 (2009).

\bibitem{gk09} Y. Gao and M.L. Kilfoil, Intermittent and spatially heterogeneous single-particle dynamics close to colloidal gelation, Phys. Rev. E {\bf 79}, 051406 (2009).

\bibitem{fhv16} E. Fodor, H. Hayakawa, P. Visco and F. van Wijland, Active cage model of glassy dynamics, Phys. Rev. E {\bf 94}, 012610 (2016).
 
\bibitem{ris07} P.M. Reis, R.A. Ingale and M.D. Shattuck, Caging dynamics in a granular fluid, Phys. Rev. Lett. \textbf{98}, 188301 (2007). 

\end{thebibliography}
\end{document}